\documentclass{aipproc}

\layoutstyle{6x9}

\usepackage{amssymb,array,stmaryrd,xypic}
\usepackage[all]{xy}
\usepackage[T1]{fontenc}
\usepackage[english]{babel} 
\usepackage[latin1]{inputenc}

\def\a{\alpha}
\def\b{\beta}

\def\Ga{\Gamma}
\def\d{\delta}

\def\k{\kappa}
\def\l{\lambda}

\def\m{\mu}
\def\om{\omega}
\def\Om{\Omega}

\newcommand{\bbR}{\mathbb{R}}

\newcommand{\bbC}{\mathbb{C}}

\newcommand{\bbN}{\mathbb{N}}

\newcommand{\bbQ}{\mathbb{Q}}

\newcommand{\bi}{\mathsf{i}}

\newcommand{\cA}{{\mathcal{A}}}

\newcommand{\tC}{C_{\cT}}
\newcommand{\Cd}{C_{\cS}}

\newcommand{\Cinfty}{{\mathcal{C}^{\infty}}}

\newcommand{\bbCl}{\mathrm{\bbC l}}

\newcommand{\sD}{\mathsf{D}}
\newcommand{\D}{\mathcal{D}}

\newcommand{\Dslm}{\mathsf{D^{\l,\m}}}

\newcommand{\Div}{\mathrm{Div}}

\newcommand{\cF}{{\mathcal{F}}}

\newcommand{\fkg}{{\mathfrak{g}}}

\newcommand{\tg}{\tilde{\gamma}}

\newcommand{\calH}{\mathcal{H}}
\newcommand{\cJ}{\mathcal{J}}

\newcommand{\ccL}{\mathcal{L}}
\newcommand{\sL}{\mathsf{L}}
\newcommand{\LD}{\mathcal{L}_X^{\lambda,\mu}}

\newcommand{\sLl}{\mathsf{L}_X^\lambda}
\newcommand{\sLm}{\mathsf{L}_X^\mu}
\newcommand{\Ld}{L_X^\delta}

\newcommand{\ro}{\mathrm{o}}

\newcommand{\Pol}{\mathrm{Pol}}

\newcommand{\QG}{{\mathcal{Q}}}
\newcommand{\cQ}{{\mathcal{Q}}}
\newcommand{\cQlm}{\mathcal{Q}^{\l,\m}}

\newcommand{\cM}{{\mathcal{M}}}

\newcommand{\cS}{{\mathcal{S}}}
\newcommand{\rS}{\mathrm{S}}
\newcommand{\rST}{\mathrm{S}_{\cT}}
\newcommand{\sS}{\mathsf{S}}

\newcommand{\cT}{\mathcal{T}}

\newcommand{\Vect}{\mathrm{Vect}}

\newcommand{\tp}{\tilde{p}}
\newcommand{\txi}{\tilde{\xi}}

\newcommand{\half}{\frac{1}{2}}

\begin{document}

\baselineskip=18pt

\newtheorem{thm}{Theorem}%[section]
\newtheorem{lem}[thm]{Lemma}
\newtheorem{cor}[thm]{Corollary}
\newtheorem{prop}[thm]{Proposition}
\newtheorem{defi}[thm]{Definition}
\newtheorem{ex}[thm]{Example}
\newtheorem{rmk}[thm]{Remark}

%%%%%%%%%%%%%%%%%%%%%%%%%%%%%%%%%%%%%%%%%%%%%%%%%%%%%%%%%%%%
%%%%%%%%%%%%%%%%%%%%%%%%% TITLE %%%%%%%%%%%%%%%%%%%%%%%%%%%%
\title{Equivariant quantization of spin systems}
%%%%%%%%%%%%%%%%%%%%%%%%%%%%%%%%%%%%%%%%%%%%%%%%%%%%%%%%%%%%
%%%%%%%%%%%%%%%%%%%%%%%%%%%%%%%%%%%%%%%%%%%%%%%%%%%%%%%%%%%%

\author{
Jean-Philippe~Michel\footnote{I thank the Luxembourgian NRF for support via the AFR grant PDR-09-063.}}{address={Institut Camille Jordan, Universit\'e Claude Bernard Lyon $1$, $43$ boulevard du $11$ novembre $1918$, F-$69622$ Villeurbanne Cedex France }\\ and\\{University of Luxembourg, Campus Kirchberg, Mathematics Research Unit, 6, rue Richard Coudenhove-Kalergi, L-1359 Luxembourg City, Grand Duchy of Luxembourg. }
}

\date{\today}


\thispagestyle{empty}

\begin{abstract}
We investigate the geometric and conformally equivariant quantizations of the supercotangent bundle of a pseudo-Riemannian manifold $(M,g)$, which is a model for the phase space of a classical spin particle. This is a short review of our previous works \cite{Mic09,Mic10a}.
\end{abstract}

\maketitle

%\vskip1cm
\noindent

%%%%%%%%%%%%%%%%%%%%%%%%%%%%%%%%%%%%%%%%%%%%%%%%%%%%%%%%%%%%
%%%%%%%%%%%%%%%%%%%%%%%%%%%%%%%%%%%%%%%%%%%%%%%%%%%%%%%%%%%%
\section{Introduction}
%%%%%%%%%%%%%%%%%%%%%%%%%%%%%%%%%%%%%%%%%%%%%%%%%%%%%%%%%%%%
%%%%%%%%%%%%%%%%%%%%%%%%%%%%%%%%%%%%%%%%%%%%%%%%%%%%%%%%%%%%

Quantization is born with quantum mechanics, as the fundamental attempt to establish a correspondence between the mathematical structures of classical and quantum mechanics, presented in the following table.

\begin{center}
  \begin{tabular}{c|c|c}
  & classical & quantum\\
  \hline
Phase space & symplectic manifold  $(\cM,\om)$ & Hilbert space $\calH$\\
\hline
Observables & Poisson algebra  $A\subset\Cinfty(\cM)$ & associative algebra $\cA\subset\ccL(\calH)$\\
\hline
Symmetries & Lie subalgebra $\fkg\subset\mathrm{ham}(\cM,\om)$ & Lie subalgebra $\fkg\subset\mathrm{u}(\calH)$. 
  \end{tabular}\\
  \end{center}
%  \vspace{0.5cm}
One of the most celebrated quantization procedure is the geometric quantization \cite{Kos70,Sou70}, whose main draw-back is its too small set of quantizable observables. Equivariant quantization \cite{LOv99,DLO99} aims to overcome this issue for systems admitting a configuration space $M$ with a large enough group $G$ of (local) symmetries, as the projective or conformal group. More precisely, the inverse of the obtained quantization map is a $G$-equivariant symbol map on $M$, from differential operators to symmetric tensors. 

We present here the equivariant quantization of spin systems whose configuration space is a spin manifold $M$ endowed with a metric $g$ of signature $(p,q)$.   This suppose to introduce a framework for classical mechanics of spin systems, namely the supercotangent bundle of $(M,g)$ \cite{BMa77,Get83} endowed with its canonical symplectic structure \cite{Rot90}. We recover some of the main objects of spin geometry via its geometric quantization, and the privileged status of conformal transformations of $(M,g)$ is highlighted. Restricting us to a conformally flat manifold, we describe explicitly the action of $\ro(p+1,q+1)$ on the space of classical and quantum observables, plus on a related space of tensors. We state then our main results on the existence and uniqueness of conformally equvariant quantization and superization, which are isomorphisms between these three $\ro(p+1,q+1)$-modules. Some applications are given. We refer to \cite{Mic09,Mic10a} for more details and proofs.

%%%%%%%%%%%%%%%%%%%%%%%%%%%%%%%%%%%%%%%%%%%%%%%%%%%%%%%%%%%%
%%%%%%%%%%%%%%%%%%%%%%%%%%%%%%%%%%%%%%%%%%%%%%%%%%%%%%%%%%%%
\section{From classical to quantum spin systems}
%%%%%%%%%%%%%%%%%%%%%%%%%%%%%%%%%%%%%%%%%%%%%%%%%%%%%%%%%%%%
%%%%%%%%%%%%%%%%%%%%%%%%%%%%%%%%%%%%%%%%%%%%%%%%%%%%%%%%%%%%
The quantum framework for spin systems with configuration space $(M,g)$ is well-known: the state space $\calH$ is obtained by completion of the space of sections of the spinor bundle $\sS\rightarrow M$, and we choose the algebra $\D(M,\sS)$ of spinor differential operators as space of quantum observables. Its usual algebra of symbols is $\Pol(T^*M)\otimes\Ga(\bbCl(M,g))$, the tensor product of the space of functions on $T^*M$, which are polynomial in the fiber variables, with the space of sections of the complex Clifford bundle of $(M,g)$. Replacing $\Ga(\bbCl(M,g))$ by its graded counterpart, namely the algebra of complex differential forms $\Om_\bbC(M)$, we end up with a superalgebra of functions on the supercotangent bundle of $M$. That provides us with the algebra of symbols for $\D(M,\sS)$ w.r.t. its bifiltration \cite{Get83}, as well as with the classical setting for a spin system on $M$.

%%%%%%%%%%%%%%%%%%%%%%%%%%%%%%%%%%%%%%%%%%%%%%%%%%%%%%%%%%%%
\subsection{Supercotangent bundle and pseudomechanics}
%%%%%%%%%%%%%%%%%%%%%%%%%%%%%%%%%%%%%%%%%%%%%%%%%%%%%%%%%%%%   
The supercotangent bundle of the manifold $M$ is $\cM=T^*M\oplus\Pi TM$, i.e. the direct sum of the cotangent bundle and the tangent bundle with reverse parity. Thus, its superalgebra of functions is $\Cinfty(T^*M)\otimes\Om(M)$, generated locally by coordinates $(x^i,p_i,\xi^i)$, where $\xi^i$ identifies with $dx^i$. The general study of symplectic supermanifolds by Rothstein \cite{Rot90} proves that a symplectic structure on $\cM$ is equivalent to the data of a metric and a compatible connexion on $M$. As a consequence, to any pseudo-Riemannian manifold $(M,g)$ corresponds a canonical symplectic form $\om$ on $\cM$, given by
\begin{equation}
\om=d\a \quad \text{and} \quad \a=p_idx^i+\frac{\hbar}{2\bi}g_{ij}\xi^id^\nabla\xi^j,
\end{equation} 
where $d^\nabla$ is the covariant differential w.r.t. the Levi-Civita connexion of $g$. The spin components are defined as $S^{ij}=\frac{\hbar}{\bi}\xi^i\xi^j$ so that, together with the Poisson bracket associated to $\om$, they generate a Lie algebra isomorphic to $\ro(p,q)$. The equations of motion of a spin particle in an exterior electromagnetic field can be easily recovered in that framework \cite{BMa77,Mic09}, and also the coupling of the spin with the gravitation \cite{Rav80,Mic09}. Thus, the Hamiltonian flows of the kinetic energy $g^{ij}p_ip_j$ leads to the Papapetrou's equations \cite{Pap51},
\begin{eqnarray*}
&&\dot{x}^j\nabla_j\dot{x}^i=-\half g^{ik}R(S)_{jk}\dot{x}^j,\\ 
&&\dot{x}^k\nabla_k S^{ij}=0,
\end{eqnarray*}
where the spin happens to be coupled with the curvature via $R(S)_{jk}=g_{im}R^i_{ljk}S^{jk}$, with $(R^i_{ljk})$ the components of the Riemann tensor.

%%%%%%%%%%%%%%%%%%%%%%%%%%%%%%%%%%%%%%%%%%%%%%%%%%%%%%%%%%%%
\subsection{Spin geometry via geometric quantization}
%%%%%%%%%%%%%%%%%%%%%%%%%%%%%%%%%%%%%%%%%%%%%%%%%%%%%%%%%%%%   

Thanks to geometric quantization, we can built the main objects of spin geometry from $(\cM,\om)$ endowed with a polarization, i.e. a Lagrangian distribution. Upon topological restrictions on $M$, the vertical polarization of the cotangent bundle of $M$ can be completed by a maximal isotropic complex distribution for $g$ on $\Pi TM$, to give a polarization on $(\cM,\om)$. In the simplest case of a Riemannian metric, the geometric quantization leads then to the construction of Hitchin \cite{Hit74} for the spinor bundle $\sS$ of a pseudo-Hermitian manifold $(M,g)$, where spinor fields identify with antiholomorphic differential forms tensorized with square root of the volume form of $(M,g)$. 

Besides, for Darboux coordinates $(x^i,\tp_i,\txi^i)$ of $(\cM,\om)$, the quantum map\footnote{we consider here that $\QG$ takes its values in usual spinor fields rather than spinor half-densities.} satisfies  
\begin{equation}
\QG(x^i)=x^i,\qquad\QG(\tp_i)=\frac{\hbar}{\bi}\partial_i \quad \text{and}\quad \QG(\txi^i)=\frac{\tg^i}{\sqrt{2}},
\end{equation} 
where $\tg^i$ is a Clifford matrix for the flat metric given by $(\eta_{ij})=\mathbb{I}_p\oplus-\mathbb{I}_q$. Let us remind that the vector fields on $M$ can be lifted to Hamiltonian vector fields on $T^*M$, giving rise to a tautological momentum map $J$, such that $J_X=p_iX^i$ for $X=X^i\partial_i\in\Vect(M)$. This map can be lifted to $\cM$ and quantized,
\begin{equation}
\QG(J_X)=\frac{\hbar}{\bi}\nabla_X,
\end{equation}
giving rise to an essential object of the spin geometry: the covariant derivative of spinors.

%%%%%%%%%%%%%%%%%%%%%%%%%%%%%%%%%%%%%%%%%%%%%%%%%%%%%%%%%%%%
\subsection{Conformal geometry of the supercotangent and spinor bundles}
%%%%%%%%%%%%%%%%%%%%%%%%%%%%%%%%%%%%%%%%%%%%%%%%%%%%%%%%%%%%   

The symplectic structure of the supercotangent bundle $\cM$ depends on a metric $g$ on~$M$, as a consequence the natural lift to $\cM$ of $X\in\Vect(M)$ does not preserve the potential $1$-form $\a$. Moreover, the condition $L_{\tilde{X}}\a=0$ does not uniquely determine the lift $\tilde{X}$ of~$X$. One way out is to impose the further condition $L_{\tilde{X}}\beta\propto\beta$, where $d\b=g_{ij}d^\nabla\xi^i\wedge dx^j$ is the odd symplectic form on $\Pi TM$. The lift is then unique but exists only if $X$ is a conformal Killing vector field. We refer to it later as $\tilde{X}$.
Let us notice that, as expected, the momentum of an infinitesimal rotation $X_{ij}$ is, 
\begin{equation}
\cJ_{X_{ij}}=p_ix_j-x_jp_i+S_{ij}.
\end{equation}
Remarkably, the quantization of the momentum map $\cJ$ leads to
\begin{equation}
\QG(\cJ_X)=\frac{\hbar}{\bi}\sL_X,
\end{equation} 
the Lie derivative of spinors introduced by Kosmann \cite{Kos72}, well-defined precisely for conformal Killing vector fields. Via geometric quantization, we get thus the conformal geometry of the spinor bundle of $(M,g)$ out of the one of its supercotangent bundle.   

%%%%%%%%%%%%%%%%%%%%%%%%%%%%%%%%%%%%%%%%%%%%%%%%%%%%%%%%%%%%
%%%%%%%%%%%%%%%%%%%%%%%%%%%%%%%%%%%%%%%%%%%%%%%%%%%%%%%%%%%%
\section{Conformally equivariant quantization}
%%%%%%%%%%%%%%%%%%%%%%%%%%%%%%%%%%%%%%%%%%%%%%%%%%%%%%%%%%%%
%%%%%%%%%%%%%%%%%%%%%%%%%%%%%%%%%%%%%%%%%%%%%%%%%%%%%%%%%%%%

We suppose from now on that $(M,g)$ is a conformally flat spin manifold, i.e. $g_{ij}=F\eta_{ij}$ for some positive function $F$. The conformal Killing vector fields on $(M,g)$ generate then a Lie algebra isomorphic to $\ro(p+1,q+1)$, which is the one of infinitesimal conformal transformations of $(\bbR^n,\eta)$. The aim is to compare the action of those vector fields on $\sD(M,\sS)$, the algebra of spinor differential operators, with those on its algebras of symbols, i.e. to compare their $\ro(p+1,q+1)$-module structures. 

%%%%%%%%%%%%%%%%%%%%%%%%%%%%%%%%%%%%%%%%%%%%%%%%%%%%%%%%%%%%
\subsection{Conformal geometry of spinor differential operators and of their symbols}
%%%%%%%%%%%%%%%%%%%%%%%%%%%%%%%%%%%%%%%%%%%%%%%%%%%%%%%%%%%%
Let us define the space of $\l$-densities by $\cF^\l=\Ga(|\Lambda^n T^*M|^{\otimes\l})$, with $\l\in\bbR$ and $n=\dim M$.   
Instead of $\sD(M,\sS)$, we will rather study the two parameters family of $\ro(p+1,q+1)$-modules $(\Dslm)$. Each of those modules is defined as the space of differential operators $D:\Ga(\sS)\otimes\cF^\l\rightarrow\Ga(\sS)\otimes\cF^\m$, endowed with the adjoint action 
\begin{equation}
\LD D=\sLl D-\sLm D,
\end{equation}
where $\sLl=\sL_X+\l\Div(X)$ is the action of $X$ on $\Ga(\sS)\otimes\cF^\l$. The corresponding ${\ro(p+1,q+1)}$-module of classical observables is naturally the space $\cS^\d[\xi]=\Pol(T^*M)\otimes\Om_\bbC(M)\otimes\cF^\d$, with $\d=\mu-\l$, endowed with the Hamiltonian action 
\begin{equation}
\Ld =\tilde{X}+\d\Div(X).
\end{equation}
The explicit expressions of these both classical and quantum actions have been computed in \cite{Mic10a}, showing that $\Dslm$ and $\cS^\d[\xi]$ are filtered modules, by the order of differential operators and the degree in the $p$ variables respectively. Their common associated graded module is the module of tensorial symbols $\cT^\d[\xi]=\bigoplus_{\k=0}^n\Pol(T^*M)\otimes\Omega^{\k}_\bbC\otimes\cF^{\d-\frac{\k}{n}}$, endowed with the natural action on weighted tensors. It identifies to $\Pol(T^*M)\otimes\Om_\bbC(M)$ as an algebra and to the usual space of symbols $\Pol(T^*M)\otimes\Ga(\bbCl(M,g))$ as a module.

%%%%%%%%%%%%%%%%%%%%%%%%%%%%%%%%%%%%%%%%%%%%%%%%%%%%%%%%%%%%
\subsection{Main results}
%%%%%%%%%%%%%%%%%%%%%%%%%%%%%%%%%%%%%%%%%%%%%%%%%%%%%%%%%%%%   
Let us begin with few definitions. A map is called conformally equivariant if it is an isomorphism of $\ro(p+1,q+1)$-modules. Besides, as $\Pol(T^*M)$ is a submodule of $\cT^0[\xi]$,  a linear isomorphism $\cT^\d[\xi]\rightarrow\cS^\d[\xi]$ which preserves the principal symbol\footnote{that is the higher order term of an element w.r.t. a filtration, here it identifies to a tensor over $M$.} is named a superization, whereas a linear isomorphism $\cS^\d[\xi]\rightarrow \Dslm$ preserving the principal symbol is a quantization, since it relates classical and quantum observables. 
 
\begin{thm}
There exists $ I^e_{\rS}\subset I_{\rS}\subset\bbQ_+^*$ such that the conformally equivariant superization 
$
\rST^\d:\cT^\delta[\xi]\rightarrow \cS^\delta[\xi],
$
exists if $\d\not\in I^e_{\rS}$ and is unique if $\d\not\in I_{\rS}$. 
\end{thm}   
\begin{thm}
Let $\d=\mu-\l\in\bbR$. There exists $I^e_{\cQ}\subset I_{\cQ}\subset\bbQ_+^*$ such that the conformally equivariant quantization 
$
\cQlm:\cS^\delta[\xi]\rightarrow \Dslm,
$
exists and is unique if $\d\not\in I_{\cQ}$, and exists for at least one value of $\l\in\bbR$ if $\d\not\in I^e_{\cQ}$.
\end{thm}   
The values of $\d$ for which existence or uniqueness of $\cQlm$ is lost are called resonances. The fact that $\d=0$ is not a resonance is crucial, thus the conformally equivariant quantization $\cQ^{\half,\half}$  extends uniquely the quantization map provided by geometric quantization.  
Let us remark that $\cQlm\circ\rST^\d$ is a particular case of AHS-equivariant quantization \cite{CSi09}, but each single map deserves interest, at least from a physical point of view.

The idea of the proofs is the same than in the spinless case \cite{DLO99}, and relies on the use of the Casimir operators of each module. To be concrete, we name $\tC$ and $\Cd$ those of $\cT^\d[\xi]$ and $\cS^\d[\xi]$. If the conformally equivariant superization exists, then $\Cd\rST^\d=\rST^\d\tC$, and in particular every eigenvectors of $\tC$ is send to an eigenvector of $\Cd$ with the same eigenvalue and the same principal symbol. The main point is to prove that the eigenvectors of $\tC$ are uniquely determined by their eigenvalue and their principal symbol. Then, if this is the case for those of $\Cd$ too, we get the uniqueness of the superization. As $\Cd$ is equal to $\tC$ plus an operator lowering the degree in the $p$ variable, this is simply checked by the resolution of triangular systems, and the resonant values of $\d$ are precisely those leading to degenerated systems. 
If unique, the map constructed in this way is easily proved to be conformally equivariant.

%%%%%%%%%%%%%%%%%%%%%%%%%%%%%%%%%%%%%%%%%%%%%%%%%%%%%%%%%%%%
\subsection{Some applications}
%%%%%%%%%%%%%%%%%%%%%%%%%%%%%%%%%%%%%%%%%%%%%%%%%%%%%%%%%%%%   

We give now two applications that we hope to investigate further in forthcoming papers. The first one relies on the explicit formulas for the conformally equivariant superization that we determine in \cite{Mic09} for symbols of degree $1$ in $p$. Let us recall that a (conformal) Killing-Yano tensor on $M$ is a skew-symmetric tensor describing higher symmetries of $(M,g)$. As Killing tensors, it generates constant of motion but for spin particles  \cite{GRV93}. In fact, a correspondence has been obtained in \cite{Tan95} between Killing-Yano tensors and classical supercharges, which happens to be generalized by the conformally equivariant superization. 
 \begin{thm}
Let $f$ be a skew-symmetric tensor and $P_f=f^i_{j_1\ldots j_{\k-1}}\xi^{j_1}\ldots\xi^{j_{\k-1}}p_i$ the associated tensor symbol. Denoting by $\Delta=p_i\xi^i$, we have
\begin{equation}
\{\Delta,\rST^0(P_f)\}=0\;(\propto\Delta ) \Longleftrightarrow f \text{ is a (conformal) Killing-Yano tensor}.
\end{equation}
\end{thm}

The second application deals with the conformal invariants of the $\ro(p+1,q+1)$-modules that we have introduced. Let $R=g^{ij}p_ip_j$. The Weyl theory of invariants together with the explicit actions of $\ro(p+1,q+1)$ on each module leads to the following Theorem.
\begin{thm}
The conformal invariants of each family of modules are
\begin{enumerate}
\item $\Delta^a\, R^s \in\cT^{\frac{2s+a}{n}}[\xi],\quad$ where $s\in\bbN$ and $a=0,1$. 
\item $\Delta\, R^s \in\cS^{\frac{2s+a}{n}}[\xi],\quad$ where $s\in\bbN$. 
\item $\cQlm(\Delta\, R^s) \in\sD^{\frac{n-2s-1}{2n},\frac{n+2s+1}{2n}},\quad$ where $s\in\bbN$. 
\end{enumerate}
\end{thm}
In particular, we get the Dirac operator as the lower order conformal invariant of $(\Dslm)$. Since the maps $\rST^\d$ and $\cQlm$ preserve conformal invariance,  we deduce from the last Theorem that $\rST^\frac{2s}{n}$ does not exist and that the module $\sD^{\frac{n-2s-1}{2n},\frac{n+2s+1}{2n}}$ is exceptional: this the only one in the family $(\sD^{\l,\l+\frac{2s+1}{n}})$ to be isomorphic to its space of symbols $\cS^\frac{2s+1}{n}[\xi]$. As a consequence, $\d=\frac{2s+1}{n}$ are resonances. In fact, every resonances correspond to some conformal invariants, that will be the matter of our next paper.

 \subsection*{Acknowledgements}
 It is a pleasure to acknowledge Christian Duval for his essential guidance, and invaluable discussions. Special thanks are due to Valentin Ovsienko for his constant interest in this work.

% ==================================================================
% BIBLIOGRAPHIE
% Gestion am�lior�e de la bibliographie
\bibliographystyle{plain}
\bibliography{BiblioWGMP}

\begin{thebibliography}{10}

\bibitem{BMa77}
F.~A. {Berezin} and M.~S. {Marinov}.
\newblock Particle spin dynamics as the grassmann variant of classical
  mechanics.
\newblock {\em Annals of Physics}, 104:336--362, April 1977.

\bibitem{CSi09}
A.~Cap and J.~Silhan.
\newblock Equivariant quantizations for {AHS}-structures.
\newblock {\em Advances in Mathematics}, 224(4):1717 -- 1734, 2010.

\bibitem{DLO99}
C.~Duval, P.~B.~A. Lecomte, and V.~Yu. Ovsienko.
\newblock Conformally equivariant quantization: existence and uniqueness.
\newblock {\em Ann. Inst. Fourier (Grenoble)}, 49(6):1999--2029, 1999.

\bibitem{Get83}
E.~Getzler.
\newblock Pseudodifferential operators on supermanifolds and the
  {A}tiyah-{S}inger index theorem.
\newblock {\em Comm. Math. Phys.}, 92(2):163--178, 1983.

\bibitem{GRV93}
G.~W. Gibbons, R.~H. Rietdijk, and J.~W. {van Holten}.
\newblock S{USY} in the sky.
\newblock {\em Nuclear Phys. B}, 404(1-2):42--64, 1993.

\bibitem{Hit74}
N.~Hitchin.
\newblock Harmonic spinors.
\newblock {\em Advances in Math.}, 14:1--55, 1974.

\bibitem{Kos72}
Y.~Kosmann.
\newblock D{\'e}riv{\'e}es de {L}ie des spineurs.
\newblock {\em Ann. Mat. Pura Appl. (4)}, 91:317--395, 1972.

\bibitem{Kos70}
B.~Kostant.
\newblock Quantization and unitary representations. {I}. {P}requantization.
\newblock In {\em Lectures in modern analysis and applications, {III}}, pages
  87--208. Lecture Notes in Math., Vol. 170. Springer, Berlin, 1970.

\bibitem{LOv99}
P.~B.~A. Lecomte and V.~Yu. Ovsienko.
\newblock Projectively equivariant symbol calculus.
\newblock {\em Lett. Math. Phys.}, 49(3):173--196, 1999.

\bibitem{Mic09}
J.-P. Michel.
\newblock {\em Quantification conform{\'e}ment {\'e}quivariante des fibr{\'e}s
  supercotangents}.
\newblock PhD thesis, Universit{\'e} Aix-Marseille II, 2009.
\newblock Electronically available as tel-00425576.

\bibitem{Mic10a}
J.-P. Michel.
\newblock Conformal geometry of the supercotangent and spinor bundles.
\newblock {\em arXiv:1004.1595 [math.DG]}, 2010.

\bibitem{Pap51}
A.~Papapetrou.
\newblock Spinning test-particles in general relativity. {I}.
\newblock {\em Proc. Roy. Soc. London. Ser. A.}, 209:248--258, 1951.

\bibitem{Rav80}
F.~Ravndal.
\newblock Supersymmetric {D}irac particles in external fields.
\newblock {\em Phys. Rev. D (3)}, 21(10):2823--2832, 1980.

\bibitem{Rot90}
M.~Rothstein.
\newblock The structure of supersymplectic supermanifolds.
\newblock In {\em Differential geometric methods in theoretical physics
  ({R}apallo, 1990)}, volume 375 of {\em Lecture Notes in Phys.}, pages
  331--343. Springer, Berlin, 1991.

\bibitem{Sou70}
J.-M. Souriau.
\newblock {\em Structure des syst{\`e}mes dynamiques}.
\newblock Ma{\^i}trise de math{\'e}matiques. Dunod, Paris, 1970 ($\copyright$
  1969).

\bibitem{Tan95}
M.~Tanimoto.
\newblock The role of {K}illing-{Y}ano tensors in supersymmetric mechanics on a
  curved manifold.
\newblock {\em Nuclear Phys. B}, 442(3):549--560, 1995.

\end{thebibliography}
 
\end{document}